 \documentclass{emulateapj}

\usepackage{graphicx,natbib}
\usepackage{hyperref}

\newcommand{\m}{$^{\rm m}\!\!.$}

\shorttitle{Ten Kepler eclipsing binaries containing the third components.} \shortauthors{Zasche et
al.}

\begin{document}

\title{Ten Kepler eclipsing binaries containing the third components.}

\author{P. Zasche\altaffilmark{1}, M. Wolf\altaffilmark{1}, H. Ku\v{c}\'akov\'a\altaffilmark{1},
J. Vra\v{s}til\altaffilmark{1}, J. Jury\v{s}ek\altaffilmark{1,2}, M. Ma\v{s}ek\altaffilmark{2}, and
M. Jel\'{\i}nek\altaffilmark{3}}
 \affil{
 \altaffilmark{1} Astronomical Institute, Charles University in Prague, Faculty of
Mathematics and Physics, CZ-180 00 Praha 8, V Hole\v{s}ovi\v{c}k\'ach 2, Czech Republic \\
 \altaffilmark{2} Institute of Physics of the Czech Academy of Sciences, CZ-182~21 Praha 8,  Na Slovance 1999/2,
Czech Republic \\
 \altaffilmark{3} Instituto de Astrof\'{\i}sica de Andaluc\'{\i}a, P.O. Box 03004, E-18080 Granada, Spain
 }

\begin{abstract}
\noindent Analyzing the available photometry from the Kepler satellite and other databases, we
performed the detailed light curve modelling of ten eclipsing binary systems, which were found to
exhibit a periodic modulation of their orbital periods.
  All of the selected systems are detached ones of Algol-type, having the orbital periods from
0.9 to 2.9 days. In total, 9448 times of minima for these binaries were analysed, trying to
identify the period variations caused by the third bodies in these systems. The well-known method
of the light-travel time effect was used for the analysis.
  The orbital periods of the outer bodies were found to be between 1 and 14 years. This hypothesis
makes such systems interesting for a future prospective detection of these components, despite
their low predicted masses. Considering the dynamical interaction between the orbits, the most
interesting seems to be the system KIC~3440230, where one would expect detection of some effects
(i.e. changing the inclination) even after a few years or decades of observations.

\end{abstract}


\section{Introduction}

The eclipsing binaries (EBs) provide us with an excellent method how to derive the basic physical
properties of the two eclipsing components (their radii, masses, temperatures). Moreover, they can
also serve as independent distance indicators, one can study the dynamical evolution of the orbits,
test the stellar structure models, or discover additional components in these systems (see e.g.
\citealt{2006Ap&SS.304....5G}). On the other hand, the Kepler satellite \citep{2010Sci...327..977B}
provides us with unprecedented accuracy of photometric data. From this huge set of observations,
1879 eclipsing binaries were detected after the first data release \citep{2011AJ....141...83P},
later extended to 2165 \citep{2011AJ....142..160S}.

Such a huge database of eclipsing binaries observed with superb precision and monitored
continuously over a period of four years encouraged several teams for looking for a periodic
modulation of data, indicating the triple systems. The use of such method and its limitations were
described elsewhere (e.g. \citealt{Irwin1959}, or \citealt{Mayer1990}). For example
\cite{2012AJ....143..137G} presented 41 suspected triples, while \cite{2014AJ....147...45C} listed
236 potential triples. More is also promised to be published by J.A.Orosz (see
\citealt{2014AJ....147...45C}), but it was not published yet. Moreover, \cite{2013ApJ...768...33R}
presented 39 dynamically interesting systems, where the third-body periods are short enough (if
compared with the binary period), hence some interaction between the orbits is expected or even
observed (e.g. changing of the inclination). On the other hand, most of the triples listed in
\cite{2014AJ....147...45C} have periods of the order of hundreds or even thousands of days. So long
periods were usually only estimated (due to limited coverage of the Kepler data), or are influenced
by large errors.

From this reason, we decided to perform a similar analysis of detecting the third-body signals for
some other systems, but based on a larger data set, if available. For some of the systems, we tried
to observe additional ground-based observations. These were done quite recently, hence even a
single point can help us to better constrain the third-body period. And finally, we also tried to
find some photometry from other sources, like the survey data from SuperWASP
\citep{2006PASP..118.1407P}, NSVS \citep{2004AJ....127.2436W}, ASAS \citep{2002AcA....52..397P},
and others. These (mostly rather scattered) points help us to prove a long-term stability of the
orbital period of the close pair, or its evolution (e.g. the quadratic ephemeris).

\section{Selection process for the binaries}

All the studied systems were chosen according to their remarkable variations in the $O-C$ diagrams.
Such ten systems naturally complete a set of triple systems as presented by
\cite{2012AJ....143..137G} and \cite{2014AJ....147...45C}. However, these two published studies
presented only such binaries, in which the third body variations are visible on the Kepler data
set, and the ones with longer periodic modulation were omitted or only briefly mentioned. This is
the main impact of the present paper. We decided to study also these systems, where the orbital
periods of the third bodies are longer and we harvested for such an analysis also the ground-based
surveys and our new photometric data. Obviously, this also leads to the conclusion that the
multiplicity fraction should be even higher than resulted from the previous studies, because a
non-negligible number of triples has the third-body orbital period of the order of years, decades,
or even longer.

For the systems under our analysis we have chosen only such systems which fulfill the following
criteria: \textcircled{\footnotesize 1} All of them are Algol-type detached binaries with circular
orbits. This information was taken from the visual inspection of the Kepler eclipsing binary
catalogue\footnote{http://keplerebs.villanova.edu/}. \textcircled{\footnotesize 2} All have
remarkable curvatures in their $O-C$~diagrams, which was considered on the basis of
\cite{2012AJ....143..137G} and \cite{2014AJ....147...45C} minima times plotted into the $O-C$
diagrams in the $O-C$ gateway\footnote{http://var.astro.cz/ocgate/} \citep{2006OEJV...23...13P}.
\textcircled{\footnotesize 3} None of these systems was studied before concerning the third-body
orbits (only a brief remark in \cite{2012AJ....143..137G} with no orbital solution is not counted
for). \textcircled{\footnotesize 4} For each of them also some additional photometry exists (older
or a more recent one) besides the Kepler data. At this point it is worth to mention that two of the
analysed systems (KIC~7630658, and KIC~9007918) were not included in the previous work on Kepler
triples detected by eclipse timing by \cite{2012AJ....143..137G}. Therefore, we have to emphasize
that due to these rather limited selection mechanisms our study does not aim to present a complete
sample for any statistical analysis of Kepler EBs. As a by-product some systems were found to
exhibit no visible variation or yielded rather spurious results yet, see below.

\section{Photometry and light curve modelling}

The analysis of the light curves (hereafter LC) based on the Kepler photometry was carried out
using the program {\sc PHOEBE} \citep{Prsa2005} for all of the systems. This program is
based on the Wilson-Devinney algorithm \citep{1971ApJ...166..605W} and its later modifications.
However, some of the parameters have to be fixed during the fitting process. The limb darkening
coefficients were interpolated from the van Hamme's tables \citep{vanHamme1993}. The albedo
coefficients $A_i$, the gravity darkening coefficients $g_i$, and also the synchronicity parameters
$F_i$ were also computed during the fitting process due to the high quality of the photometry.
The same apply for the value of the third light, which was also considered as a free
parameters and has been fitted (in agreement with our third-body hypothesis). The temperature of
the primary component was kept fixed according to the $T_1$ value as given in the Kepler
catalogue\footnote{http://archive.stsci.edu/kepler/data\underbar{ }search/search.php}
\citep{2011AJ....142..112B}, while only the secondary temperature was fitted. In our final
solution we only present a ratio of the temperatures $T_2/T_1$ for a higher robustness due to
(sometimes) problematic values of the $T_1$ from the Kepler catalogue. An issue of the mass ratio
was solved by fixing $q=1$ because no spectroscopy for these selected systems exists, and for
detached eclipsing binaries the LC solution is almost insensitive to the photometric mass ratio
(see e.g. \citealt{2005Ap&SS.296..221T}).

\begin{figure}
 \includegraphics[width=0.48\textwidth]{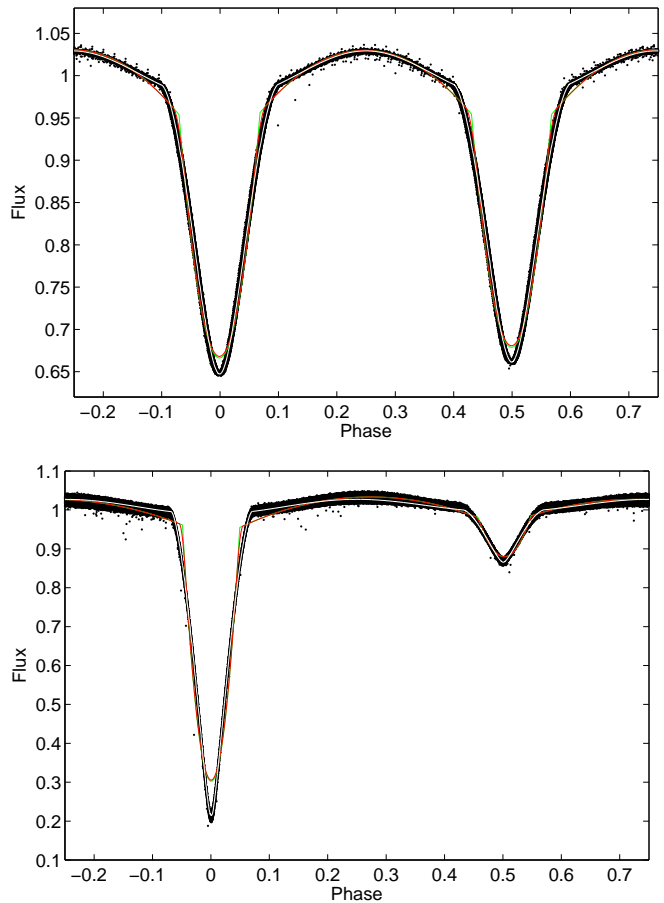}
 \caption[]{Illustrative example of the light curves of KIC~09402652 (upper plot), and KIC~10581918
 (lower plot) together with their respective fits -- from \cite{2011AJ....142..160S} plotted in red
 and green, and our new fit plotted in white.} \label{FigsLC}
\end{figure}

The quality of the LC fit is even noticeable by a naked eye, see Fig. \ref{FigsLC}. The automatic
routines as used e.g. by \cite{2011AJ....142..160S} are definitely better for reduction of a huge
data sets of hundreds of binaries, however the codes sometimes produce spurious results. If we
analyse the particular system in more detail, we are able to get a better fit to the data, lower
residuals, and hence also parameters with lower errors. The automatic pipelines maybe even not
compute parameters like albedos, gravity brightening, third light, etc. All of these parameters can
be also fitted with the Kepler data and help us to obtain a better fit to the data.

However, it is necessary to admit that some of the parameters can correlate with each other during
the fitting process (this especially apply e.g. for luminosity and temperature, inclination and the
third light for partially eclipsing systems, etc). This problem was avoided checking whether there
is some value in the correlation matrix higher than 0.8 and such a fit was not accepted. Another
iteration with different parameter set was used and this way all the systems were analysed yielding
the results presented below.

\section{The times of minimum light}

The CCD follow-up observations of selected Kepler targets were mostly carried out in Ond\v{r}ejov
Observatory in the Czech Republic (labelled as OND in the tables with minima times), few
new observations were also obtained remotely with the BOOTES-1A and
BOOTES-2\footnote{http://bootes.iaa.es/} telescopes located in Spain (labelled as BOO-1,
and BOO-2 in the tables with minima times). The new times of primary and secondary minima and
their respective errors were determined by the classical \cite{1956BAN....12..327K} method or by
our new approach (see below). All the times of minima used for the analysis are given in the
appendix Tables \ref{minima}.

For the analysis of minima times and variation of orbital period caused by the third body one needs
the minima times as precise as possible. For some of the Kepler targets the times of minima exist
and were even published several times, see e.g. \cite{2012AJ....143..137G}, or
\cite{2014AJ....147...45C}. However, their published times of minima differ significantly --
sometimes more than their respective errors. One possible explanation is that the above-mentioned
teams included and not included an error in barycentric times of the Kepler data, which was first
mentioned in "Kepler Data Release 19"\footnote{{ \fontsize{2}{2}\selectfont
https://archive.stsci.edu/kepler/release\underbar{ }notes/release\underbar{
}notes19/DataRelease\underbar{ }19\underbar{ }20130204.pdf}}.

Due to this reason, we have proceeded in the following way. At first, the times of minima published
by \cite{2012AJ....143..137G} were taken and plotted into the $O-C$~diagram. The same was done with
the data by \cite{2014AJ....147...45C} and the $O-C$ diagrams were analysed whether some periodic
modulation due to the third body is presented. Then, the original data from the Kepler archive were
downloaded and analysed. Such an analysis was done in several steps. At first, from the original
raw \texttt{fits} files the photometry was extracted, the flux converted into magnitudes and the
individual light curves in different quarters of data were analysed and the theoretical light
curves were constructed. Lets call this way the Method 1. On the other hand, the Method 2 was using
the data downloaded from the EB catalogue\footnote{http://keplerebs.villanova.edu/}  by
\cite{2011AJ....142..160S}, which were already detrended and the normalised flux versus BJD was
provided. These light curves were analysed and used to construct the theoretical light curve (but
from the whole Kepler mission).

The theoretical light curves were used to derive the times of minima following the AFP method as
described in \cite{2014A&A...572A..71Z}. Using the LC templates from Method 1 and 2 and also the
times of minima from \cite{2012AJ....143..137G} and \cite{2014AJ....147...45C}, we have four
different sets of times of minima for the analysis (disadvantage of the \cite{2012AJ....143..137G}
data is the fact that only a portion of the Kepler data was provided, these ones after reducing the
first 9 quarters only). These four data sets differ significantly sometimes and the best one (with
the lowest scatter) was used for the subsequent analysis of a particular system. Usually, the best
one was the data set obtained by Method 2.

However, it is natural that some limitations of the method play a role. The most critical issue is
the fact that for deriving the times of minima we always use the same LC template. However, for
some cases the shape of the LC varies during the Kepler mission and the difference is sometimes
visible even by naked eye (see below comments for particular systems). This problem can be avoided
using the different LC templates for data obtained during the different time epochs. However, it is
a questionable task whether using five or a hundred different LC templates for the whole Kepler
data set would provide a better result. Hence, we solved out this problem by using a
slightly different template for each Kepler quarter.

If we compare both minima derivation methods, we found some aspects of the problem. The
classical \citeauthor{1956BAN....12..327K} method was used only for recent observations due to the
fact that only small parts of the minima were observed and the whole LC cannot be fitted. On the
other hand, the AFP method can provide us with much more precise result even with lower number of
observations, but one needs the complete LC template, hence the complete observed LC. Generally,
the individual errors from the AFP method are a bit lower (but not 10 times lower) than the
classical errors from the \citeauthor{1956BAN....12..327K} method and are not affected by any
observational biasses, wrong reduction, poor conditions, etc. as can be true for the ground-based
ones.

\section{The period changes} \label{PerChange}

For the analysis of period changes in these binaries, we used a well-known method introduced by
\cite{Irwin1959}. It resulted in a set of parameters of the third-body orbit: period of the third
body $P_3$, eccentricity $e$, semi-amplitude of the variation $A$, time of periastron passage
$T_0$, and the longitude of periastron $\omega$. The input values for the analysis were the
ephemerides ($HJD_0$, $P$) given by \cite{2011AJ....142..160S}, while also these ephemerides were
recomputed. If necessary, also the quadratic term of the ephemerides was used (attributed to the
mass transfer between the components). The solutions presented below were found using Monte
Carlo simulations and the simplex algorithm. However, the individual errors of parameters are taken
from the code and may be too optimistic for some of the systems.

All the new precise CCD times of minima from the Kepler satellite were used with a weight of 10 in
our computation; some of the less precise measurements were weighted by a factor of five, while the
poorly covered minima were given a weight of 1. This apply mostly for the minima times derived from
other sources of photometry (like ASAS, SuperWASP, etc.), which were derived using the same method
as the Kepler ones, but using a different LC template. The weights were used instead of the
uncertainties due to the fact that for the older published minima any information about their
accuracy is missing.

Because of studying only the period changes due to the third-body orbit, and all of the systems are
circular, for most of the systems only the deeper (primary) minimum was used to detect the period
changes.


\begin{table*}[h!]
\caption{Relevant information for the analysed systems.}  \label{InfoSystems}
 \scriptsize
\begin{tabular}{lccccccccccc}
   \hline\hline\noalign{\smallskip}
  System      & Other ID                &          RA           &             DE                     &$KEP_{\rm max}^{A}$&$(J-H)[\mathrm{mag}]^{B}$ & $(B-V)[\mathrm{mag}]^{C}$& Sp.Type$^C$ \\
  \hline\noalign{\smallskip}
 KIC  2305372 & 2MASS J19275768+3740219 & 19$^h$27$^m$57$^s$.7 & +37$^\circ$40$^\prime$21$^{\prime\prime}$.9 &  13\m82       & 0.364  &       &       \\ 
 KIC  3440230 & 2MASS J19215310+3831428 & 19$^h$21$^m$53$^s$.1 & +38$^\circ$31$^\prime$42$^{\prime\prime}$.8 &  13\m64       & 0.317  &       &       \\ 
 KIC  5513861 & TYC 3123-2012-1         & 18$^h$57$^m$24$^s$.5 & +40$^\circ$42$^\prime$52$^{\prime\prime}$.9 &  11\m64       & 0.238  & 0.448 & wF8V  \\ 
 KIC  5621294 & 2MASS J19285262+4053359 & 19$^h$28$^m$52$^s$.6 & +40$^\circ$53$^\prime$36$^{\prime\prime}$.0 &  13\m61       & 0.143  &       &       \\ 
 KIC  7630658 & 2MASS J19513965+4315224 & 19$^h$51$^m$39$^s$.6 & +43$^\circ$15$^\prime$22$^{\prime\prime}$.3 &  13\m89       & 0.389  &       &       \\ 
 KIC  8553788 & 2MASS J19174291+4438290 & 19$^h$17$^m$42$^s$.9 & +44$^\circ$38$^\prime$29$^{\prime\prime}$.1 &  12\m69       & 0.120  & 0.537 & A7V   \\ 
 KIC  9007918 & TYC 3541-2296-1         & 19$^h$04$^m$02$^s$.0 & +45$^\circ$21$^\prime$21$^{\prime\prime}$.7 &  11\m66       & 0.135  & 0.155 & F5IV  \\ 
 KIC  9402652 & V2281 Cyg               & 19$^h$25$^m$06$^s$.9 & +45$^\circ$56$^\prime$03$^{\prime\prime}$.1 &  11\m82       & 0.154  & 0.470 & F8V   \\ 
 KIC 10581918 & WX Dra                  & 18$^h$52$^m$10$^s$.5 & +47$^\circ$48$^\prime$16$^{\prime\prime}$.7 &  12\m80       & 0.186  &       &       \\ 
 KIC 10686876 & TYC 3562-961-1          & 19$^h$56$^m$13$^s$.6 & +47$^\circ$54$^\prime$33$^{\prime\prime}$.7 &  11\m73       &(-0.041)& 0.204 & F0V   \\ 
 \noalign{\smallskip}\hline
\end{tabular}
 \scriptsize Note: [A] - Kepler database, [B] - 2MASS catalogue, \cite{2006AJ....131.1163S}, [C] -
 based on the Tycho catalogue, \cite{2010PASP..122.1437P}.
\end{table*}

\begin{table*}[h!]
\caption{Light curve parameters for the analysed systems as resulted from the {\sc PHOEBE}.}
\label{LCparam}
 \scriptsize
\begin{tabular}{lcccccccccc}
  \hline\hline\noalign{\smallskip}
   System     &   $T_2/T1$         &  $i$ [deg]   &  $\Omega_1$   &  $\Omega_2$   &  $L_1$ [\%]  &  $L_2$  [\%] & $L_3$ [\%] \\
  \hline\noalign{\smallskip}
 KIC  2305372 &  0.6637 (0.0152) & 79.92 (0.27) & 5.431 (0.035) & 4.134 (0.059) & 82.70 (0.90) & 17.30 (0.80) &  0 \\
 KIC  3440230 &  0.6082 (0.0085) & 81.63 (0.82) & 6.278 (0.692) & 5.114 (0.192) & 87.02 (0.83) & 12.98 (0.47) &  0 \\
 KIC  5513861 &  0.9891 (0.0115) & 79.37 (0.08) & 5.393 (0.012) & 5.773 (0.024) & 55.10 (0.23) & 43.97 (0.27) &  0.94 (0.55) \\
 KIC  5621294 &  0.5620 (0.0096) & 72.32 (0.73) & 4.182 (0.084) & 4.255 (0.106) & 82.85 (0.35) &  8.86 (3.02) & 11.29 (0.99) \\
 KIC  7630658 &  0.9635 (0.0004) & 79.76 (0.02) & 7.660 (0.005) & 7.646 (0.007) & 51.66 (0.02) & 43.26 (0.02) &  5.07 (0.02) \\
 KIC  8553788 &  0.6385 (0.0022) & 69.72 (0.22) & 5.351 (0.025) & 5.106 (0.057) & 80.15 (0.71) & 13.27 (0.15) &  6.56 (0.60) \\
 KIC  9007918 &  0.6289 (0.0008) & 72.83 (0.05) & 5.479 (0.006) & 5.781 (0.016) & 79.06 (0.05) &  6.36 (0.02) & 14.58 (0.05) \\
 KIC  9402652 &  0.9956 (0.0033) & 79.61 (0.07) & 4.386 (0.007) & 4.357 (0.004) & 50.01 (1.68) & 49.99 (1.44) &  0 \\
 KIC 10581918 &  0.6813 (0.0126) & 88.53 (0.42) & 5.595 (0.058) & 5.751 (0.050) & 86.68 (0.67) & 13.32 (0.50) &  0 \\
 KIC 10686876 &  0.6532 (0.0048) & 88.35 (0.06) & 6.976 (0.030) &16.290 (0.123) & 92.32 (3.41) &  2.76 (0.11) & 4.92 (3.08) \\
 \noalign{\smallskip}\hline
\end{tabular}
\end{table*}

\begin{table*}[h!]
\caption{The parameters of the third-body orbits for the individual systems.} \label{OCparam}
 \tiny
\begin{tabular}{ccccccccccc}
\hline\hline\noalign{\smallskip}
   System     & $HJD_0        $  &   $P$         &  $A$         & $\omega$       & $P_3$      & $T_0$ [HJD]   &      e     & $f(m_3)$   & $P_3^2/P$  \\   
              & (2450000+)       &   [days]      & [days]       &  [deg]         & [yr]       & (2400000+)    &            & $[M_\odot]$& [yr]    \\
 \noalign{\smallskip}\hline\noalign{\smallskip}
 KIC  2305372 & 4965.9539 (8)  & 1.4047173 (15) & 0.0211 (13)  &   86.9 (4.7)   &10.36 (0.16)& 54532 (62)  & 0.625 (66) & 0.4543 (18) & 27919  \\  
 KIC  3440230 & 5687.5150 (3)  & 2.8811052 (38) & 0.00060 (25) &  111.3 (17.5)  & 1.04 (0.13)& 55818 (32)  & 0.264 (98) & 0.0010 (1)  &   137  \\  
 KIC  5513861 & 4955.0004 (9)  & 1.5102096 (10) & 0.00831 (73) &   27.2  (7.4)  & 5.94 (0.18)& 56347 (139) & 0.135 (89) & 0.0861 (39) &  8540  \\  
 KIC  5621294 & 4954.5109 (2)  & 0.9389102 (3)  & 0.00024 (5)  &  133.9 (11.7)  & 2.70 (0.10)& 56124 (28)  & 0.654 (175)& 0.000014 (2)&  2843  \\  
 KIC  7630658 & 5003.2780 (2)  & 2.1511554 (4)  & 0.00393 (3)  &  145.9 (1.1)   & 2.53 (0.01)& 67358 (17)  & 0.680 (10) & 0.0875 (33) &  1085  \\  
 KIC  8553788 & 4954.9856 (13) & 1.6061776 (17) & 0.00802 (114)&  237.3 (8.6)   & 9.09 (0.08)& 56430 (59)  & 0.764 (93) & 0.0429 (50) & 18787  \\  
 KIC  9007918 & 4954.7485 (2)  & 1.3872066 (2)  & 0.00048 (4)  &   94.6 (9.2)   & 1.30 (0.08)& 56721 (14)  & 0.662 (170)& 0.00034 (4) &   445  \\  
 KIC  9402652 & 4954.2856 (2)  & 1.0731067 (2)  & 0.00427 (17) &  266.9 (4.9)   & 4.08 (0.05)& 56343 (11)  & 0.757 (37) & 0.0242 (10) &  5670  \\  
 KIC 10581918 & 2829.3696 (3)  & 1.8018668 (34) & 0.00209 (58) &    0.1 (8.3)   &14.05 (0.47)& 53244 (530) & 0.254 (88) & 0.00027 (9) & 40023  \\  
 KIC 10686876 & 4953.9490 (45) & 2.6184137 (50) & 0.00563 (189)&  280.5 (28.4)  & 6.72 (0.96)& 56990 (442) & 0.464 (157)& 0.0207 (19) &  6302  \\  
 \noalign{\smallskip}\hline
\end{tabular}
\end{table*}

\section{The individual systems}

In the following section we present the results of our analysis for all of the systems. The whole
procedure is described in detail for the first binary, the others are only briefly discussed due to
similarity of the analysis with the first one. The Table \ref{InfoSystems} summarizes basic
information about the stars, their cross-identification, magnitudes and photometric indices. As one
can see from the $(J-H)$ index, most of the stars are of F and G spectral type.

\subsection{KIC 2305372}

The first system in our sample is the star KIC 2305372, which was first recognized as a variable by
Hatnet \citep{2004AJ....128.1761H} and ASAS \citep{2009AcA....59...33P} surveys in the pre-Kepler
era. After then, it was included into the catalogue of eclipsing binaries in the Kepler field
\citep{2011AJ....142..160S}. The times of minima were published by \cite{2012AJ....143..137G} and
later by \cite{2014AJ....147...45C}. However, \cite{2012AJ....143..137G} presented the system as a
candidate triple, while \cite{2014AJ....147...45C} roughly estimated some period of about 3700
days. No spectral analysis was carried out, hence we can only estimate that it is probably a system
of G spectral type (from the $J-H$ photometric index).

\begin{figure*}[h!]
  \includegraphics[width=0.85\textwidth]{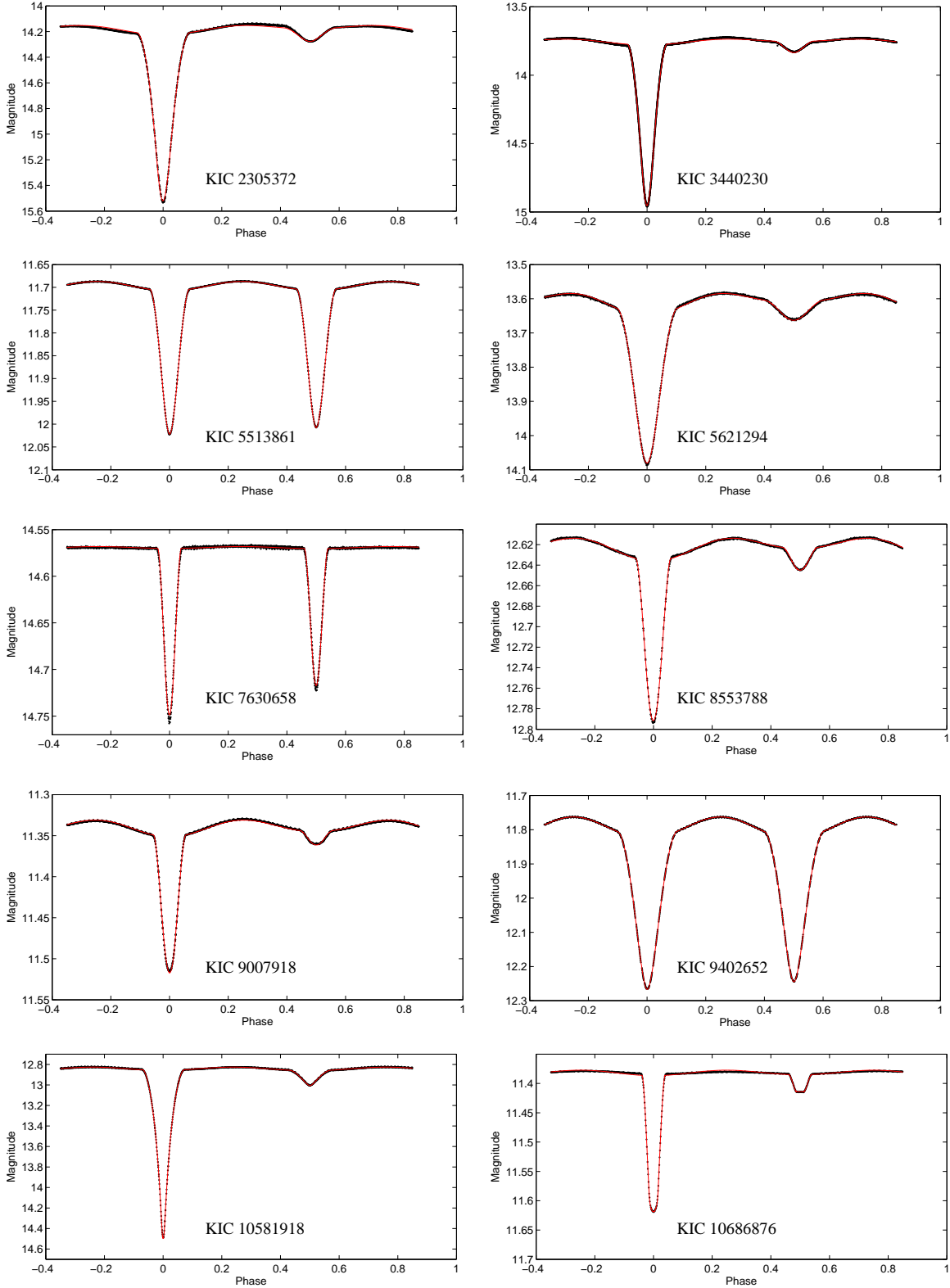}
 \caption[]{The Kepler light curves of all studied systems. The red curves present the final fit,
 the dots stand for the observations.} \label{figLC}
\end{figure*}

The light curve analysis was carried out from the Kepler detrended data, while its parameters are
given in Table \ref{LCparam}. As one can see, both components are rather different, while no third
light was detected during the LC solution. The final LC fit is presented in Fig.\ref{figLC}, where
is clearly seen a shape of the LC as a classical Algol one. However, the LC shape seems to be
slightly asymmetric (see the outside eclipse curvature). This light curve template was also used
for deriving the times of minima (using the method as described above). For the period analysis we
collected the Hatnet, ASAS, SuperWASP and the Kepler data points and derived more than 800 times of
primary minima for this star. One new minimum was also observed by the authors at Ond\v{r}ejov
Observatory in the Czech Republic.

\begin{figure*}[h!]
 \includegraphics[width=0.9\textwidth]{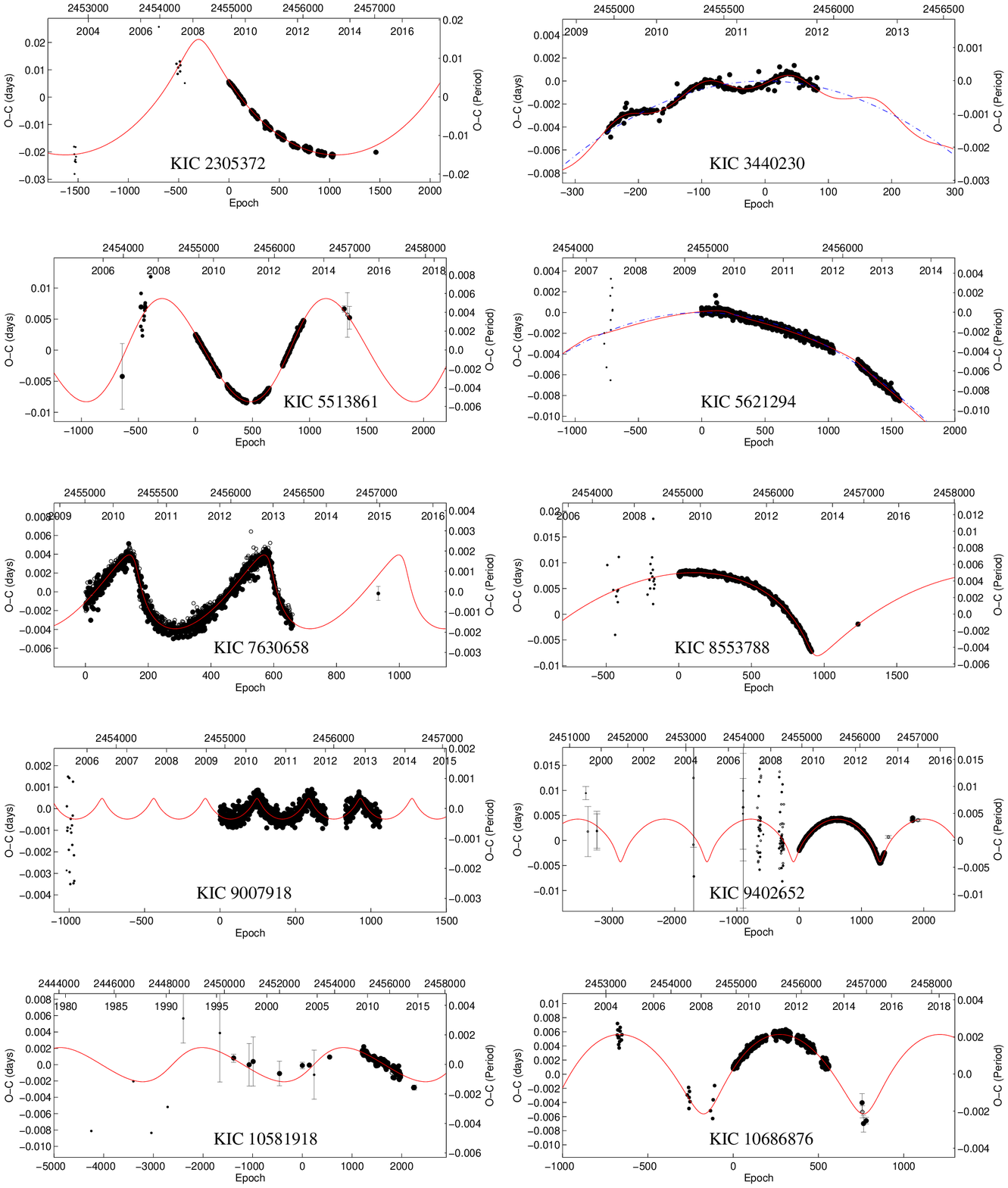}
 \caption[]{The $O-C$ diagrams of all studied systems. The red curves present the final fit, the
 blue dash-dotted curves the quadratic ephemerides. The dots stand for the primary minima,
 open circles for the secondary minima, while the bigger the symbol, the higher the weight in our
 computation (the oldest visual observations are plotted as small dots and their respective errors
 were even not published, but we estimate their precision to be up to 5--15 minutes).} \label{figOC}
\end{figure*}

This data set was analysed and the method of \cite{Irwin1959} was used. The results are given in
Table \ref{OCparam} and the final fit is plotted also in the Figure \ref{figOC}. In these plots
only the new post-Kepler data and the isolated measurements (groups of up to three data points) are
plotted with their respective error bars for a better clarity. Plotting the error bars for all the
data would diminish the readability of the graphs (however, for some observations their respective
error bars are too small and are plotted almost inside the individual dots). We are aware of the
fact that only a few points of poor quality define the shape of the third-body variation and its
period $P_3$ in the $O-C$ diagram. However, the parabolic fit is not able to describe the data in
such detail. From the parameters of the third body one is able to compute also the mass function of
the third body in the system, which is also given in Table \ref{OCparam}. As one can see, its value
is rather high, so the third component should be detected also in the LC solution as a third light
contribution. However, no such value was detected during the LC fitting. This still remains an open
question, however we also have to mention that the shape of the LC varies in time and the LC fit in
different quarters of data differs a bit. This can also influence our result and the minima
precision, LC modelling and the third light detection. Regrettably, having no information about the
masses of the eclipsing components, one cannot easily set a tighter limit to the mass of the
predicted third body.

\subsection{KIC 3440230}

KIC 3440230 was discovered by \cite{2011AJ....142..160S}, later \cite{2012AJ....143..137G} included
the star into the group of tertiary candidates. On the other hand, there was also a remark about
the flux variation and possible pulsations \citep{2012AJ....143..137G}. This is the star with the
longest orbital period in our sample.

The same method as for the previous star was used. We were not able to fit the outside-eclipse
curvature of the Kepler LC (due to asymmetry of the LC), but the primary minimum is fitted pretty
well. Therefore, the LC template was used for deriving the minima times used for a subsequent
period analysis. Besides the Kepler data also a few SuperWASP minima were derived. However, these
were not used for the analysis due to their large scatter. The long-term period decrease is also
visible on the Kepler data with no need to spread the time interval with these scattered data
points. From the third-body orbit fitting there resulted a very small mass function value
$$f(m_3) = \frac {(m_3 \sin i)^3} {(m_1+m_2+m_3)^2} = \frac {1}{P_3^2} \cdot {\left[ \frac {173.15
\cdot A} {\sqrt{1-e^2\cos^2\omega}} \right]^3},$$ which is mostly caused by the small amplitude of
the variation. The potential third body would probably be of a late-type dwarf star.

On the other hand, what makes this system the most interesting is the fact that the period $P_3$ is
rather short, hence one can hope to detect some dynamical interaction between the orbits (see e.g.
\citealt{2013ApJ...768...33R}, \citealt{1975A&A....42..229S}). The nodal period can be computed
from the equation
$$P_{\mathrm{nodal}} = \frac{4}{3} \left( 1+ \frac{m_1 + m_2}{m_3} \right) \frac{P_3^2}{P} (1-e_3^2)^{3/2} \left( \frac{C}{G_2} \cos j \right)^{-1},$$
where the subscripts 1 and 2 stand for the eclipsing binary components, while 3 stands for the
third distant body, the term ${G_2}$ stands for the angular momentum of the wide orbit, the $C$ is
the total angular momentum of the system, and $j$ stands for the mutual inclination of the
orbits. For this system the ratio of periods $P_3^2/P$ resulted in surprisingly low value of about
137~yr only. Hence, one can hope to detect some changes of the binary orbit even after a few years
of observations. The most promising is the inclination change, because it is rather easily
detectable. Due to its deep eclipses a change of the inclination angle should be detected also in
the ground-based data of a modest quality. However, the amplitude of any such change is also
strongly dependent upon a third-body mass and orientation of its orbit. For derivation of these
quantities a precise interferometry or spectroscopy would be very useful. However, one cannot hope
to obtain these observations for a 14-magnitude star easily.

\subsection{KIC 5513861}

The star KIC 5513861 (also TYC 3123-2012-1) was first mentioned as a variable by
\cite{2009AcA....59...33P} from the ASAS data. Later, \cite{2012AJ....143..137G} reported about its
curvature in the $O-C$ diagram, probably caused by a third body. Also mentioned were the pulsations
and rapid flux variability. \cite{2014AJ....147...45C} published a preliminary results from the
Kepler data estimating that the third body should have a period of about $\approx$1800~days. This
is the first system in our sample of stars, which was included into the work by
\cite{2010PASP..122.1437P}, who used the Tycho photometry for estimating the spectral type of the
star, see Table \ref{InfoSystems}.

The same approach for the analysis was used, the LC was fitted and the final plot is then used as a
template to derive the precise times of minima. The final $O-C$ diagram is plotted in Fig.
\ref{figOC}, where also some minima as derived from the ASAS \citep{2002AcA....52..397P} and
SuperWASP \citep{2006PASP..118.1407P} surveys were included together with our three new
observations (one from Ond\v{r}ejov Observatory in the Czech Republic, two from the BOOTES-1A and
BOOTES-2 telescopes in Spain). All of these data clearly define the third-body variation with a
period of about 6 years and yielding a moderate value of the mass function. However, the fraction
of the third light is rather lower than anticipated from the third-body mass function. With the
available data we are not able to find where the problem should be and the nature of the third body
still remains an open question.

\subsection{KIC 5621294}

The system KIC 5621294 was discovered from the Kepler data \citep{2011AJ....142..160S}. Later, the
times of minima were published by \cite{2012AJ....143..137G}, who also included a remark about a
possible parabolic trend in the $O-C$ diagram, starspots and pulsations.

\begin{figure}[h!]
  \includegraphics[width=0.48\textwidth]{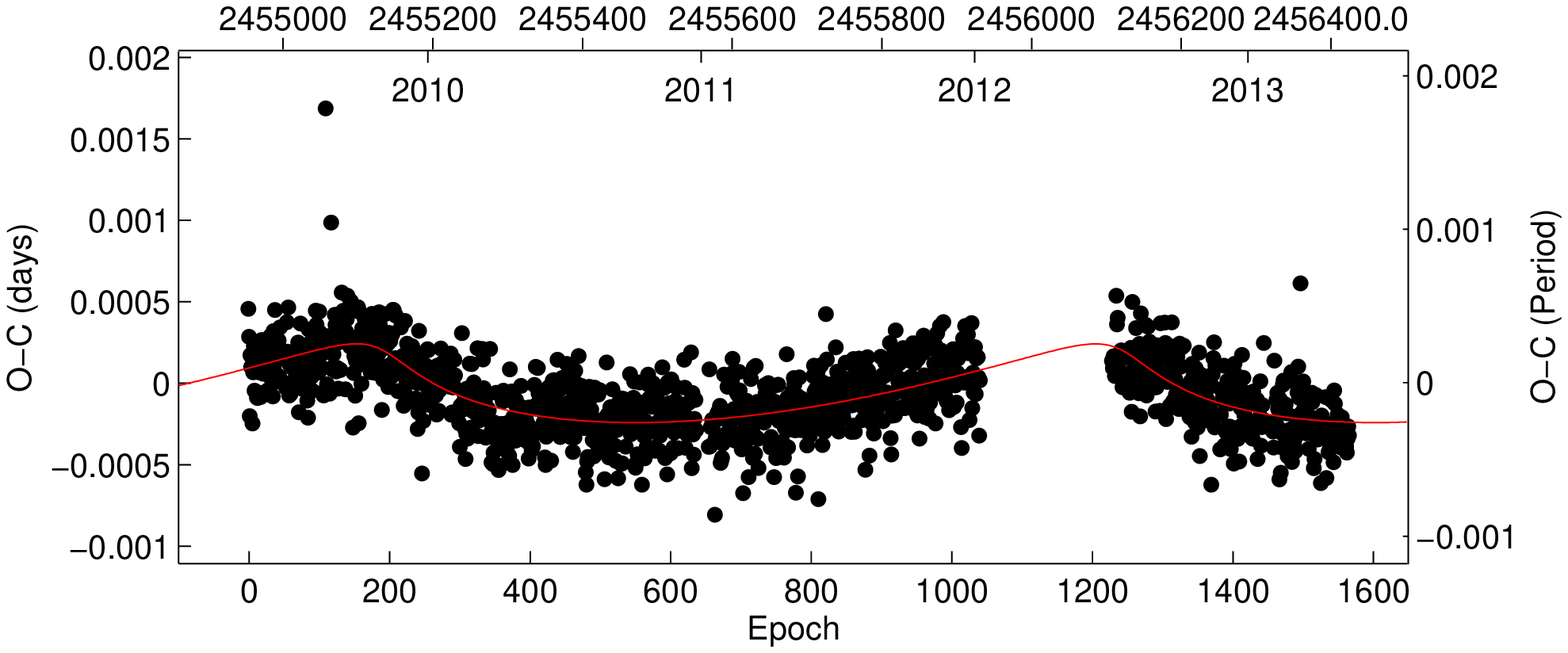}
 \caption[]{The $O-C$ diagram of the Kepler minima times of the system KIC 5621294 after subtraction of the quadratic ephemerides.} \label{figKIC5621294}
\end{figure}

The LC was fitted and analysed, resulted in the largest difference between the primary and
secondary temperatures of the eclipsing components in our sample of stars. From the LC parameters
written in Table \ref{LCparam} one can see a non-negligible value of the third light and only very
weak contribution of the secondary component to the total light. On the other hand, the times of
minima as derived from the Kepler data show a significant period decrease (described via parabolic
ephemerides), see Fig. \ref{figOC}. Moreover, superposing over the parabola also a small periodic
variation is visible with period of about 2.7~yr and the lowest amplitude in our sample (about 21
seconds only), see Fig. \ref{figKIC5621294}. This small amplitude yielded also a small value of the
predicted third light value, hence a third light contribution as detected during the LC solution
should probably be attributed to another body in the system or a close visual component.
However, it is still rather premature to speculate that we deal here with a real quadruple
system. Such a low amplitude of the variation in the $O-C$ diagram could also serve as a
testing example of what can be even discovered from the Kepler data by these classical techniques
with eclipsing binaries: assuming the component masses $M_1=M_2=1M_\odot$ then the minimum
third-body mass (i.e. assuming $i=90^\circ$) resulted in $M_{3,min}=0.039~M_\odot$, hence a typical
brown dwarf mass.

\subsection{KIC 7630658}

The system  KIC  7630658 was discovered by \cite{2011AJ....142..160S} from the Kepler data. No
other analysis was carried out and our knowledge about the system is very limited. It is the
faintest star in our sample.

The shape of the LC as obtained by the Kepler satellite clearly shows two well-defined minima,
hence also the derivation of the times of minima was rather straight-forward. The final parameters
are given in Table \ref{LCparam}, where one can see that both components are similar to each other
and only a small fraction of the third light was detected. The variation with period of about
2.5~yr is clearly visible on the data, however our last observation slightly deviates from the
prediction. This can be caused by some long-term modulation of the orbital period (quadratic
ephemerides), but this have to be tested in the upcoming years with new observations.

\subsection{KIC 8553788}

The star KIC 8553788 was first mentioned as an eclipsing binary by \cite{2009AcA....59...33P}.
Later, only the results from the Kepler data analysis were published: \cite{2011AJ....142..160S},
\cite{2011AJ....141...83P}, and \cite{2012AJ....143..137G}. The latter paper gives some information
about possible pulsations, starspots and possible third body. This system seems to be of the
earliest spectral type in our sample of stars (see Table \ref{InfoSystems}).

Our analysis using the Kepler data yielded the LC solution showing that the primary is the dominant
object in the system, hence only the primary minima were used for the $O-C$ diagram analysis. The
9-yr variation is clearly visible in the plot despite the fact that the orbital period is still
determined only by the one last observation from the Ond\v{r}ejov Observatory. The older
observations from the ASAS and SuperWASP surveys only slightly follow the predicted fit, but have
quite large scatter. Our fit of minima times yielded rather high value of eccentricity, however the
minimal third-body mass as resulted from the mass function is somewhat lower than the masses of the
eclipsing components. Its light contribution hence should probably be higher than resulted from our
LC fit.

\subsection{KIC 9007918}

The star KIC 9007918 (also TYC 3541-2296-1) was first detected as a variable by
\cite{2008AJ....135..850D} on the basis of the TRES survey data. Later, the star was included into
the catalogue of Kepler eclipsing binaries (\citealt{2011AJ....142..160S}, and
\citealt{2011AJ....141...83P}).

There were detected some variations on the LC during the Kepler mission, and the whole LC is not
perfectly symmetric. This can also play some role on the precision of the derived times of minima
from the LC template. As one can also see from the LC, the secondary minimum is only very shallow,
hence we used only the primary ones for analysing the period changes in this binary. Together with
the old (and rather scattered) photometry from the TRES survey we were able to detect the periodic
variations with the period of about 1.3~yr and an amplitude of about 41~seconds only. The other
interesting issue is also the value of period for a possible dynamical interaction between the
orbits $P_3^2/P \sim 445$~yr. Hence, we can hope to find some changes after several decades of
observations.

\subsection{KIC 9402652}

The star KIC 9402652 (also V2281 Cyg) was discovered as a variable already in the pre-Kepler era
and a few observations of the minima of this star were published. It was mentioned in the list of
stars observed by the ROTSE survey \citep{2001IBVS.5060....1D}, later \cite{2009AcA....59...33P}
included the star into their ASAS observations of the Kepler fields, and the times of minima were
published by \cite{2012AJ....143..137G} and \cite{2014AJ....147...45C}.

As one can see, the system consists of two almost identical stars, both temperatures and
luminosities are practically the same. From this reason, also both the minima are very similar,
hence both primary and secondary were used for the period analysis. We also collected the older
published minima together with the photometry from the NSVS, SuperWASP, and ASAS surveys. Thanks to
the large data set of available times of minima observations this system seems to be the richest
one in our sample of stars (and with the data coverage ranging over more than 15 years). The $O-C$
diagram together with our new observations clearly shows the 4-yr variation, but with rather high
eccentricity.

\subsection{KIC 10581918}

The system KIC 10581918 (also WX Dra) was discovered as a variable as early as in 1960 by
\cite{1960ATsir.210...22T}. Since than a few observations of the minima were published, but no
light curve nor spectroscopic analysis of the system. Due to very deep primary eclipse of this star
(1.67~mag) also the older visual and photographic observations can be reliable for the analysis of
the period changes. The very first preliminary results were published in the conference proceedings
(\citealt{Wolf2015}).

As one can see from the results of our analysis, the period of the third body is of about 14~yr
(the longest one in our sample) and is now well-covered, but its amplitude is only poorly defined
with our data. New minima times observations in the upcoming years can help us to better derive the
amplitude of variations. However, the predicted mass function of the third body resulted in rather
low value, hence also a non-detection of the third light in the LC solution is something
expectable.

\subsection{KIC 10686876}

The eclipsing binary KIC 10686876 was first mentioned by \cite{2008AJ....135..850D}, based on the
TRES survey data. Later, the star was included into the Kepler eclipsing binary database,
\cite{2011AJ....141...83P}, and \cite{2011AJ....142..160S}. \cite{2012AJ....143..137G} published
the minima times for the system, but no other information or analysis was performed.

The star seems to be the only one system in our sample which shows total eclipse. Due to
this reason also the error of the inclination from the LC fit is very small. On the other hand, the
secondary component is probably a very small star and the primary is the dominating one. As one
can also see, the primary eclipses are rather deep and provide us much better times of minima than
the secondaries. Hence, analysing the available minima from the Kepler, TRES, SuperWASP, and our
new data (two from Ond\v{r}ejov, two from the BOOTES-1A and BOOTES-2 telescopes in Spain) we
obtained a set of third-body parameters given in Table \ref{OCparam} and the final fit presented in
Fig. \ref{figOC}. The variation with a period of about 6.7~yr is now clearly visible in the current
data set and the shape of the $O-C$ variation should easily be confirmed and the parameters
improved by a few new observations obtained during the upcoming years.

\section{Discussion and conclusions}

Ten selected binaries were found to be worth of study due to the presence of the distant
components, which cause the periodic modulation of their eclipsing periods. The periods of the
third bodies (from 1 to 14 years) are usually adequately covered with the Kepler and the
ground-based data, so the variation is certain nowadays. However, its origin is still questionable
in several cases. This especially applies to such systems where the predicted mass function of the
third body and the non/detected third light from the LC solution contradict each other. However,
this can be caused by some of these reasons: 1. the imperfect LC fit (for these binaries with
slightly asymmetric LC), 2. not very well-defined third body variation in the $O-C$ diagram
(especially in these cases where the variation is mostly determined by the older scattered
ground-based data), 3. the variation in the $O-C$ diagram incorrectly described (i.e. missing
quadratic term or a fourth-body variation), 4. exotic object as the distant body (or also a binary,
hence having much lower luminosity), or 5. some other phenomena modulating the period variation in
the $O-C$ diagram (such as magnetic or other activity of the components). As a by-product of our
analysis, there were found a few more systems, where the $O-C$ variation was not found, or is still
questionable yet. These are summarized in Table \ref{OtherSystems}. Regrettably, this is still too
limited sample to do any reliable statistical analysis of incompleteness of triple systems found in
the Kepler data.

At this point it would be useful to mention that when using the "Method 1" as introduced in Section
3, some of the systems also have the short cadence data in the Kepler photometric database. Using
the short cadence produces much more precise minima derivation (these minima times are labelled as
"Kepler SC" in the Appendix table with minima), but can also reveal some other phenomena
non-detectable in the long cadence data. This happened for KIC~8553788 and KIC~10686876, for which
some short time variation was detected on the short cadence data (probably $\delta$~Sct
pulsations), which were not visible on the long cadence one. However, such additional variation
also influences the light curve fitting and its precision.

\begin{table*}
\caption{Some other analysed systems.}  \label{OtherSystems}
 \scriptsize
\begin{tabular}{lll}
   \hline\hline\noalign{\smallskip}
  System      & Other ID                & Remark  \\
  \hline\noalign{\smallskip}
 KIC 04245897 & V583 Lyr                & some variation with period about 50 yr found, but based only on older photographic data \\ 
 KIC 06187893 & TYC 3128-1653-1         & quadratic ephemerides or third body with long period, not very convincing, new data needed \\ 
 KIC 06852488 & 2MASS J19135355+4222482 & some variation detected, but period still uncertain, more data needed  \\ 
 KIC 07258889 & 2MASS J18510630+4248400 & some variation found, but showing rather non-periodic modulation  \\ 
 KIC 07938468 & V481 Lyr                & quadratic ephemerides based also on older photographic data  \\ 
 KIC 08552540 & V2277 Cyg               & no variation found \\ 
 KIC 09101279 & V1580 Cyg               & some variation found, but not very convincing, older data too scattered  \\ 
 KIC 09602595 & V0995 Cyg               & variation with period 13.3~yr found, but the data before 1970 are in contradiction \\ 
 KIC 09899416 & BR Cyg                  & no variation found  \\ 
 KIC 10736223 & V2290 Cyg               & quadratic ephemerides only, based on older visual data  \\ 
 KIC 11913071 & V2365 Cyg               & no variation found  \\ 
 KIC 12071006 & V379 Cyg                & some variation detected only on the Kepler data, older measurements too scattered  \\ 
 \noalign{\smallskip}\hline
\end{tabular}
\end{table*}

One has to consider also the limitations of the method used for the analysis. The LC fit is a
crucial issue, because it is used to derive the minima times for a subsequent analysis. However,
the LC fits can also be the problematic issue, because we are dealing with pure photometry with no
information about the individual masses of the components. Hence, fixing the mass ratio value $q=1$
is in fact only the first rough simplification. Therefore, having no information about the
individual masses, also the mass function of the third body provides only very preliminary
information about such object. Due to this reason and because of the unknown distance also the
angular separation of the third component cannot be computed for a prospective interferometric
detection. However, it should probably be hard to detect such bodies due to relative faintness of
most of the stars for this technique.

To conclude, only dedicated high-dispersion, and high-S/N spectroscopic observations and a
subsequent analysis can tell us something more about these objects and reveal their true nature.
Moreover, also some new photometric observations in the upcoming years would be of great benefit,
especially in these systems where the period variation is still not very certain yet and also for
the dynamically interesting systems like KIC~3440230.

\medskip

 \acknowledgments
 An anonymous referee is acknowledged for the useful comments and suggestions that significantly
improved the paper.
 We would like to thank Mr. Kamil Hornoch for obtaining some of the photometric
observations in the Czech Republic and A.J. Castro-Tirado for the data from BOOTES-1A and BOOTES-2
telescopes located in Spain.
 We do thank the {\sc ASAS}, {\sc Hatnet}, {\sc Kepler}, {\sc SuperWASP}, {\sc TRES} and {\sc NSVS} teams
for making all of the observations easily public available.
 This work was supported by the Czech
Science Foundation grants P209/10/0715, and GA15-02112S.
 We also acknowledge funding from the Spanish Ministry Projects AYA 2009-14000-C03-01 and
AYA2012-39727-C03-01 and the support from the BOOTES/IAA-CSIC team (R. Cunniffe, O. Lara-Gil et
al.), INTA (B. de la Morena, J. A. Adame, M. D. Sabau-Graziati et al.), UV (V. Reglero) and
IHSM/UMA-CSIC staff (R. Fern\'andez et al.).
 The operation of the robotic telescopes BOOTES-1A and BOOTES-2 are supported by the EU
grant GLORIA (No. 283783 in FP7-Capacities program) and by the grant of the Ministry of Education
of the Czech Republic (MSMT-CR LG13007).
 The following internet-based resources were used in research for this paper: the SIMBAD database
and the VizieR service operated at the CDS, Strasbourg, France, and the NASA's Astrophysics Data
System Bibliographic Services.

\begin{appendix} 

\section{Tables of minima}

\begin{table}
 \centering
  \begin{minipage}{85mm}
 \fontsize{0.95mm}{1.35mm} \selectfont
 \caption{\scriptsize List of the minima timings used for the analysis. A sample what is included in the online data repository.} \label{minima}
\begin{tabular}{ccclcl}
\hline\hline\noalign{\smallskip}
 Star       &    BJD -    &  Error  & Type   &Filter$^*$& Source /     \\
            &   2400000   &  [day]  &        &          & Observatory  \\
\noalign{\smallskip}\hline \noalign{\smallskip}
  KIC 2305372& 52802.67112 & 0.08710 & Prim   &    I     & Hatnet       \\
  KIC 2305372& 52806.88251 & 0.07516 & Prim   &    I     & Hatnet       \\
  KIC 2305372& 52809.68474 & 0.02068 & Prim   &    I     & Hatnet       \\
  KIC 2305372& 52813.90357 & 0.06025 & Prim   &    I     & Hatnet       \\
  KIC 2305372& 52816.71288 & 0.02019 & Prim   &    I     & Hatnet       \\
  KIC 2305372& 52820.92775 & 0.03789 & Prim   &    I     & Hatnet       \\
  KIC 2305372& 52823.74173 & 0.04552 & Prim   &    I     & Hatnet       \\
  KIC 2305372& 52827.95039 & 0.03124 & Prim   &    I     & Hatnet       \\
  KIC 2305372& 52830.76181 & 0.10454 & Prim   &    I     & Hatnet       \\
  KIC 2305372& 53986.89162 & 0.00559 & Prim   &    I     & ASAS         \\
  KIC 2305372& 54349.28814 & 0.00479 & Prim   &    I     & ASAS         \\
  KIC 2305372& 54232.70365 & 0.04672 & Prim   &    W     & SuperWASP    \\
  KIC 2305372& 54249.55658 & 0.21725 & Prim   &    W     & SuperWASP    \\
  KIC 2305372& 54256.58249 & 0.32321 & Prim   &    W     & SuperWASP    \\
  KIC 2305372& 54280.46490 & 0.30372 & Prim   &    W     & SuperWASP    \\
  KIC 2305372& 54284.67542 & 0.29591 & Prim   &    W     & SuperWASP    \\
  KIC 2305372& 54287.48710 & 0.08857 & Prim   &    W     & SuperWASP    \\
  KIC 2305372& 54964.55486 & 0.00138 & Prim   &    K     & Kepler       \\
  KIC 2305372& 54965.95912 & 0.00058 & Prim   &    K     & Kepler       \\
  KIC 2305372& 54967.36384 & 0.00094 & Prim   &    K     & Kepler       \\
  KIC 2305372& 54968.76843 & 0.00066 & Prim   &    K     & Kepler       \\
  KIC 2305372& 54970.17336 & 0.00101 & Prim   &    K     & Kepler       \\
  KIC 2305372& 54971.57795 & 0.00067 & Prim   &    K     & Kepler       \\
  KIC 2305372& 54972.98267 & 0.00062 & Prim   &    K     & Kepler       \\
  KIC 2305372& 54974.38726 & 0.00070 & Prim   &    K     & Kepler       \\
  KIC 2305372& 54975.79185 & 0.00079 & Prim   &    K     & Kepler       \\
  KIC 2305372& 54977.19678 & 0.00085 & Prim   &    K     & Kepler       \\
  KIC 2305372& 54978.60149 & 0.00091 & Prim   &    K     & Kepler       \\
  KIC 2305372& 54980.00574 & 0.00073 & Prim   &    K     & Kepler       \\
  KIC 2305372& 54981.41067 & 0.00086 & Prim   &    K     & Kepler       \\
  KIC 2305372& 54982.81537 & 0.00039 & Prim   &    K     & Kepler       \\
  KIC 2305372& 54984.21996 & 0.00110 & Prim   &    K     & Kepler       \\
  KIC 2305372& 54985.62488 & 0.00070 & Prim   &    K     & Kepler       \\
  KIC 2305372& 54987.02981 & 0.00062 & Prim   &    K     & Kepler       \\
  KIC 2305372& 54988.43405 & 0.00086 & Prim   &    K     & Kepler       \\
  KIC 2305372& 54989.83910 & 0.00070 & Prim   &    K     & Kepler       \\
  KIC 2305372& 54991.24321 & 0.00047 & Prim   &    K     & Kepler       \\
  KIC 2305372& 54992.64825 & 0.00082 & Prim   &    K     & Kepler       \\
  KIC 2305372& 54994.05283 & 0.00069 & Prim   &    K     & Kepler       \\
  KIC 2305372& 54995.45753 & 0.00078 & Prim   &    K     & Kepler       \\
  KIC 2305372& 54996.86211 & 0.00067 & Prim   &    K     & Kepler       \\
  KIC 2305372& 55003.88545 & 0.00040 & Prim   &    K     & Kepler       \\
  KIC 2305372& 55005.29049 & 0.00020 & Prim   &    K     & Kepler       \\
  KIC 2305372& 55006.69480 & 0.00014 & Prim   &    K     & Kepler       \\
  KIC 2305372& 55008.09956 & 0.00012 & Prim   &    K     & Kepler       \\
  KIC 2305372& 55009.50452 & 0.00014 & Prim   &    K     & Kepler       \\
  KIC 2305372& 55010.90921 & 0.00032 & Prim   &    K     & Kepler       \\
  KIC 2305372& 55012.31376 & 0.00023 & Prim   &    K     & Kepler       \\
  KIC 2305372& 55013.71867 & 0.00016 & Prim   &    K     & Kepler       \\
  KIC 2305372& 55017.93246 & 0.00021 & Prim   &    K     & Kepler       \\
  KIC 2305372& 55019.33701 & 0.00035 & Prim   &    K     & Kepler       \\
  KIC 2305372& 55020.74184 & 0.00027 & Prim   &    K     & Kepler       \\
  KIC 2305372& 55022.14659 & 0.00014 & Prim   &    K     & Kepler       \\
  KIC 2305372& 55023.55113 & 0.00026 & Prim   &    K     & Kepler       \\
  KIC 2305372& 55024.95602 & 0.00013 & Prim   &    K     & Kepler       \\
  KIC 2305372& 55026.36035 & 0.00046 & Prim   &    K     & Kepler       \\
  KIC 2305372& 55027.76524 & 0.00028 & Prim   &    K     & Kepler       \\
  KIC 2305372& 55029.16978 & 0.00028 & Prim   &    K     & Kepler       \\
  KIC 2305372& 55030.57440 & 0.00024 & Prim   &    K     & Kepler       \\
  KIC 2305372& 55031.97912 & 0.00021 & Prim   &    K     & Kepler       \\
  KIC 2305372& 55033.38386 & 0.00022 & Prim   &    K     & Kepler       \\
  KIC 2305372& 55034.78833 & 0.00034 & Prim   &    K     & Kepler       \\
  KIC 2305372& 55036.19320 & 0.00030 & Prim   &    K     & Kepler       \\
  KIC 2305372& 55037.59793 & 0.00027 & Prim   &    K     & Kepler       \\
  KIC 2305372& 55039.00247 & 0.00034 & Prim   &    K     & Kepler       \\
  KIC 2305372& 55040.40708 & 0.00019 & Prim   &    K     & Kepler       \\
  KIC 2305372& 55041.81186 & 0.00021 & Prim   &    K     & Kepler       \\
  KIC 2305372& 55043.21665 & 0.00019 & Prim   &    K     & Kepler       \\
  KIC 2305372& 55044.62138 & 0.00011 & Prim   &    K     & Kepler       \\
  KIC 2305372& 55046.02599 & 0.00028 & Prim   &    K     & Kepler       \\
  KIC 2305372& 55047.43055 & 0.00016 & Prim   &    K     & Kepler       \\
  KIC 2305372& 55048.83520 & 0.00029 & Prim   &    K     & Kepler       \\
  KIC 2305372& 55050.23952 & 0.00043 & Prim   &    K     & Kepler       \\
  KIC 2305372& 55051.64437 & 0.00034 & Prim   &    K     & Kepler       \\
  KIC 2305372& 55053.04903 & 0.00012 & Prim   &    K     & Kepler       \\
  KIC 2305372& 55054.45394 & 0.00020 & Prim   &    K     & Kepler       \\
  KIC 2305372& 55055.85838 & 0.00008 & Prim   &    K     & Kepler       \\
  KIC 2305372& 55057.26288 & 0.00050 & Prim   &    K     & Kepler       \\
  KIC 2305372& 55058.66788 & 0.00016 & Prim   &    K     & Kepler       \\
  KIC 2305372& 55060.07257 & 0.00038 & Prim   &    K     & Kepler       \\
  KIC 2305372& 55061.47687 & 0.00042 & Prim   &    K     & Kepler       \\
   \noalign{\smallskip}\hline
 \end{tabular}
 Note: * {\it W and K stand for special filters used for SuperWASP and Kepler.}
 \end{minipage}
 \end{table}

\end{appendix}

\end{document}